\documentclass[aps,prl,amssymb,showpacs,twocolumn,floatfix]{revtex4-1}
\usepackage{graphicx}
\usepackage{amsmath}
\usepackage{multirow}

\begin{document}
\title{Spin rectification for collinear and non-collinear magnetization and external magnetic field configurations }
\author{Y. Huo,$^{1, 2,}$}\email{11110190012@fudan.edu.cn}
\author{L. H. Bai,$^{2}$ P. Hyde,$^{2}$ Y. Z. Wu,$^{1}$
and
C.-M. Hu$^2$}
\affiliation{$^1$ Department of Physics, State Key Laboratory of Surface Physics and Collaborative Innovation Center of Advanced Microstructures, Fudan University, Shanghai 200433, China}
\affiliation{$^2$ Department of Physics and Astronomy, University of Manitoba, Winnipeg, Manitoba R3T 2N2, Canada }

\begin{center}
\begin{abstract}
  
Spin rectification in a single crystal Fe/Au/Fe sandwich is electrically detected for collinear and non-collinear magnetization and external magnetic field configurations. The line shape, line width and signal polarity are analysed. The spin rectification theory has been much extended by taking the magneto-crystalline anisotropy and shape anisotropy into account, which explained non-collinear resonances and agrees very well with experimental data. Thus, a comprehensive understanding of spin rectification in ferromagnetic metal was demonstrated in this work.  
  
\end{abstract}
\end{center}
\maketitle

A decade ago, spin dynamics in ferromagnetic materials was electrically detected via the spin diode effect in magneto tunnel junctions \cite{spindiode, diodeRalph} and the bolometric effect in thin films \cite{bolometric, bolometricNew}, which triggered a rapid development. Later, more methods were developed, such as the spin pumping effect \cite{Spinpumping}, the (inverse) spin hall effect \cite{Hirsch, Saitoh} and the spin rectification effect (SRE) \cite{shuji,SRE}. The SRE dominates the electrical voltage induced by ferromagnetic resonance (FMR) in a ferromagnetic metal \cite{SRETheory}. A precessing magnetization leads to a periodically changing resistance through magneto resistance. The periodically changing resistance couples with the microwave current flowing inside and generates a DC voltage, this is the SRE. Such a method became the most popular method in electrical detection of FMR because of its high sensitivity, simple sample structure, and experimental set up. It was applied to different materials and structures with accurate agreement between theory and experimental results on both line shape and line width \cite{SRETheory, SRESEP, SREFe, SREAngle,PyYIG, BaiLH, DingHF, Hoffman, FePt, BraSRE, GerSRE}. Such line shape analysis is useful for distinguishing spin rectification from spin pumping and inverse spin hall effect \cite{BaiLH, Hoffman, GerSRE}. Line width is also important for determining additional damping due to spin pumping as well as intrinsic Gilbert damping \cite{SRETheory, Spinpumping, Bret, Spinmemoryloss}. All the previous studies of line shape and line width were performed in a collinear case where the magnetization is aligned parallel with the external magnetic field. However, in ferromagnetic thin film, the magnetization orientates along an effective field direction rather than the external magnetic field direction, especially when the internal magnetic field, such as magnetic anisotropy field and demagnetization field, is comparable to the external magnetic field. In such a non-collinear case of the magnetization and the external magnetic field, the line shape and line width analyses of spin rectification haven't been systematically studied yet.

In this work, we experimentally studied the line shape and line width of spin rectification in a non-collinear case for a sample with strong anisotropy. We also extend the spin rectification theory from a collinear case into a non-collinear case by considering all anisotropy effects. Thus, we present a comprehensive understanding of spin rectification in a metallic system.

\begin{figure}[configuration]
\includegraphics[width= 7.6 cm]{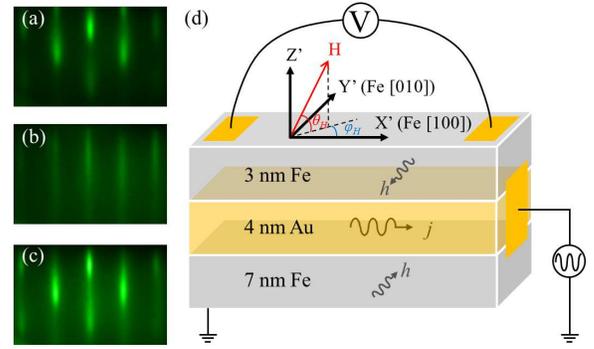}
\caption{RHEED patterns with the electron beam $e^-$ along\\ MgO $\langle$100$\rangle$ of (a) Fe (7 nm)/MgO, (b) Au (4 nm)/Fe (7 nm)\\/MgO, and (c) Fe (3 nm)/Au (4 nm)/Fe (7 nm)/MgO. (d) a sketch of measurement geometry.}
\label{Configuration}
\end{figure}

To achieve a system with strong anisotropy, we designed ultra-thin single crystal Fe/Au/Fe sandwich on MgO (001) substrate by molecular beam epitaxy in a ultra high vacuum chamber. The substrate was cleaned by annealing at 680$^{\circ}$C for 45 minutes. Then, a 7-nm-thickness Fe was prepared at room temperature and annealed at 250$^{\circ}$C for 3 minutes until the high crystalline quality achieved as indicated by a sharp reflection of high-energy electron diffraction (RHEED) pattern, as shown in Fig. 1 (a). A 4-nm-thickness Au was then epitaxially deposited at room temperature. A 3-nm-thickness of Fe was then epitaxially deposited. Further, a 5-nm-thickness MgO layer was deposited on top for protection. The RHEED patterns shown in Fig. \ref{Configuration}(a)-(c) indicate the smoothness of each layer surface and the high crystalline quality of the sample. In addition to the shape anisotropy, the single crystal Fe ultra-thin film on MgO (001) has a strong four-fold anisotropy in plane with the easy axis along the Fe [100] and the hard axis along Fe [110] \cite{FeMgO}, and the two Fe layers with different thickness perform different magnetic anisotropy field \cite{MAThickness}, which have all been confirmed in out measurement. Both magneto-crystalline anisotropy and shape anisotropy in Fe/Au/Fe sandwich allow us to study the non-collinear spin rectification in this work.

As shown in Fig. \ref{Configuration}(d), the tri-layer sample was patterned into a strip along the Fe [100] easy axis with dimension of 20 $\mu m \times $ 3 mm using standard photo lithography. A microwave was applied into the strip directly, and most microwave current flows inside of Au layer due to high conductivity. Thus, the microwave magnetic field on the bottom layer has a phase shift of $\pi$ with that in the top layer. The microwave was modulated with a frequency of 8.33 kHz. Voltage was measured along the strip using lock-in technique. An external magnetic field ${\mathbf H}$ was applied to the strip with orientation defined in Fig. 1(d). Spin rectification voltage was measured by sweeping the external magnetic field at a fixed microwave frequency. In this work, microwave power is 100 mW. Before showing our experimental results, we have ruled out the magnetization coupling \cite{Mcouple} and spin current coupling \cite{PyYIG} between two Fe layers experimentally (experimental evidence was no shown in this work). Therefore, we treat both Fe layers independently. And, we have carefully checked the special condition \cite{BaiLH} for pure spin pumping, and the signal is ignorable comparing to that of spin rectification. Thus, we were allowed to study the line shape, line width, and polarity of the pure spin rectification signal in both collinear and non-collinear cases. 
\begin{figure}[configuration]
\includegraphics[width= 7.6 cm]{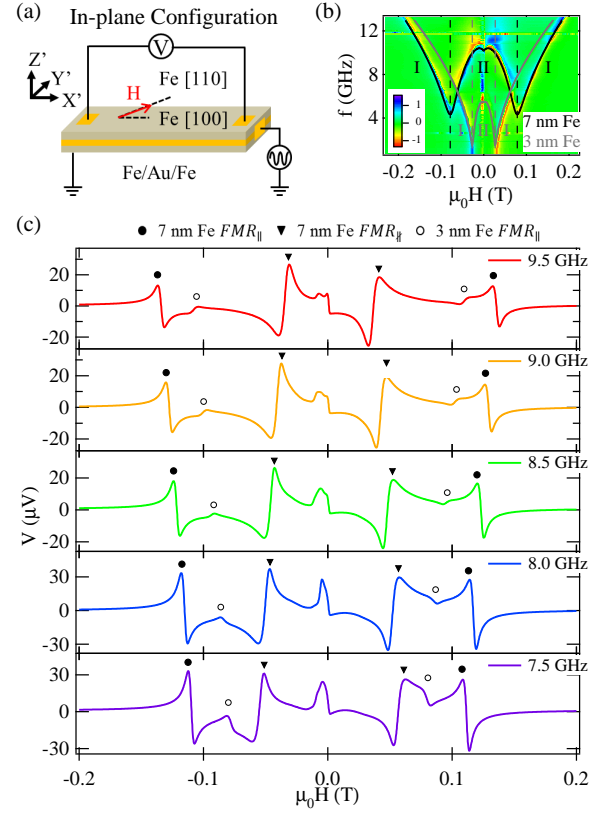}
\caption{$V_{SR}$ measurement with ${\mathbf H}$ along hard axis Fe [110] in plane. (a) The sketch of the in-plane configuration measurement; (b) $\omega$-${\mathbf H}$ dispersive image plot, the grey solid line is the fitting curve of 3 nm Fe, and the black solid line is the fitting curve of 7 nm Fe; both dispersion curves have two brunches: brunch I is $FMR_{\parallel}$ brunch, and brunch II is $FMR_{\nparallel}$ brunch. (c) Typical curves in in-plane configuration, solid circles ($\bullet$) indicate peaks belong to $FMR_{\parallel}$ brunch in 7 nm Fe, solid triangles ($\blacktriangledown$) indicate peaks  belong to $FMR_{\nparallel}$ in 7 nm Fe, hollow circles ($\circ$) indicate peaks belong to $FMR_{\parallel}$ in 3 nm Fe.}
\label{inplaneM}
\end{figure}

Figure \ref{inplaneM} shows the results when ${\mathbf H}$ is applied near Fe [110] direction in the film plane, which is the hard axis of four-fold magneto crystalline anisotropy. Fig. \ref{inplaneM}(a) shows a sketch of in-plane configuration measurement with ${\varphi _H}\approx45^\circ$ and ${\vartheta _H}=0^\circ$. For this case,  when ${\mathbf H}$ is larger than the saturation field, the magnetization ${\mathbf M}$ will lie almost parallel to the ${\mathbf H}$ direction, while if ${\mathbf H}$ is smaller than saturation field, ${\mathbf M}$ will be pulled out of the collinear configuration, and the relative angle between ${\mathbf M}$ and ${\mathbf H}$ is determined by the competition between the Zeeman energy and four-fold magneto crystalline anisotropy energy. Fig. \ref{inplaneM}(b) shows $\omega$-${\mathbf H}$ dispersion plot, with normalized rectification voltage amplitude mapped into rainbow color scale as the indicator marks. Dispersion curves are calculated by solving Landau-Lifshitz-Gilbert (LLG) equation \cite{FeMgO}, and we get four-fold magnetic anisotropy field $\mu_{0} H_{1}=0.073~T$ (black solid line) and $\mu_{0}H_{1}=0.026~T$ (grey solid line) of each Fe layer by fitting the measured data in Fig. \ref{inplaneM}(b). Due to the Fe-thickness dependence of anisotropy \cite{MAThickness}, we can identify the dispersion curve traced by the black solid line as originating from the 7 nm Fe layer and the curve traced by the grey solid line as originating from the 3 nm Fe layer. These two dispersive curves cross at $ \mu_{0}H=\pm~0.065 ~T$, and the independence of the two dispersive curves near the cross indicates the magnetic coupling between the two FM layers is very weak. Both the $\omega$ -${\mathbf H}$ dispersive curves have two brunches, as shown in Fig. \ref{inplaneM}(b). In brunch I, the resonance field increases as the frequency increases, here ${\mathbf H}$ is larger than the saturation field and thus ${\mathbf M} \parallel {\mathbf H}$; we define the resonance in this situation as $FMR_{\parallel }$ brunch. In brunch II, the resonance field decreases as the frequency increases, here ${\mathbf H}$ is smaller than the saturation field and thus ${\mathbf M} \nparallel {\mathbf H}$; we define the resonance in this situation as $FMR_{\nparallel }$ brunch. Fig. \ref{inplaneM}(c) shows some typical curves measured in this configuration at various microwave frequencies between 7.5 GHz and 9.5 GHz. All resonance in the curves are anti-symmetric Lorenz line shape, indicating the phase shift between microwave field ${\mathbf h}$ and microwave current ${\mathbf j}$ is almost the integers of $\pi$ in this device \cite{SRESEP}. In addition to the rectification voltage observed at the FMR fields of the 3 nm Fe and 7 nm Fe, a non resonant rectification signal is observed around $\mu_{0}H$=0; this signal arises due to the spin rotation which occurs as the magnetic field reverses, as discussed by X. F. Zhu, et al \cite{ZeroSRE}. In this paper we shall focus our study only on the resonance rectification voltage. From Fig. \ref{inplaneM}(c), we summarize the main features of the SRE measured in the in-plane configuration by the following Eqs. (\ref{e:V_inplane}): (a) all voltage signals change their polarity when the applied magnetic field reverses; (b) the voltage polarity in the 7 nm Fe $FMR_{\parallel}$ brunch is opposite to the polarity in the 3 nm Fe $FMR_{\parallel}$ brunch; (c) the voltage polarity in $FMR_{\parallel}$ brunch is opposite to the polarity in $FMR_{\nparallel }$ brunch.
\begin{subequations}
\label{e:V_inplane}
\begin{align}
At~{\varphi _H}\approx45^\circ, {\vartheta _H}=0^\circ:\nonumber\\
~~~~~~~~~~~~~~~~~~~~~~~~~~~~~&V(H)=-V(-H)\label{e:V_inplaneA}\\
~~~~~~~~~~~~~~~~~~~~~~~~~~~~~&\frac{V_{Fe_{7}}}{|V_{Fe_{7}}|}=-\frac{V_{Fe_{3}}}{|V_{Fe_{3}}|}\label{e:V_inplaneB}\\
~~~~~~~~~~~~~~~~~~~~~~~~~~~~~&\frac{V_{FMR_{\parallel}}}{|V_{FMR_{\parallel}}|}=-\frac{V_{FMR_{\nparallel }}}{|V_{FMR_{\nparallel }}|}\label{e:V_inplaneC}
\end{align}
\end{subequations}

Eq. (\ref{e:V_inplaneA}) is in agreement with the literatures \cite{BaiLH, SREAngle}, and Eq. (\ref{e:V_inplaneB}) describes the polarity difference in two Fe layers due to the phase shift of the microwave magnetic field. Eq. (\ref{e:V_inplaneC}) indicates that in the in-plane configuration the polarity of $V_{SR}$ changes its sign for the case where ${\mathbf M}$ and ${\mathbf H}$ are non-collinear. And from Fig. (\ref{inplaneM})(c), the resonance peaks in $FMR_{\nparallel}$ brunch is much broader than in $FMR_{\parallel}$ brunch. 

\begin{figure}[configuration]
\includegraphics[width= 7.6 cm]{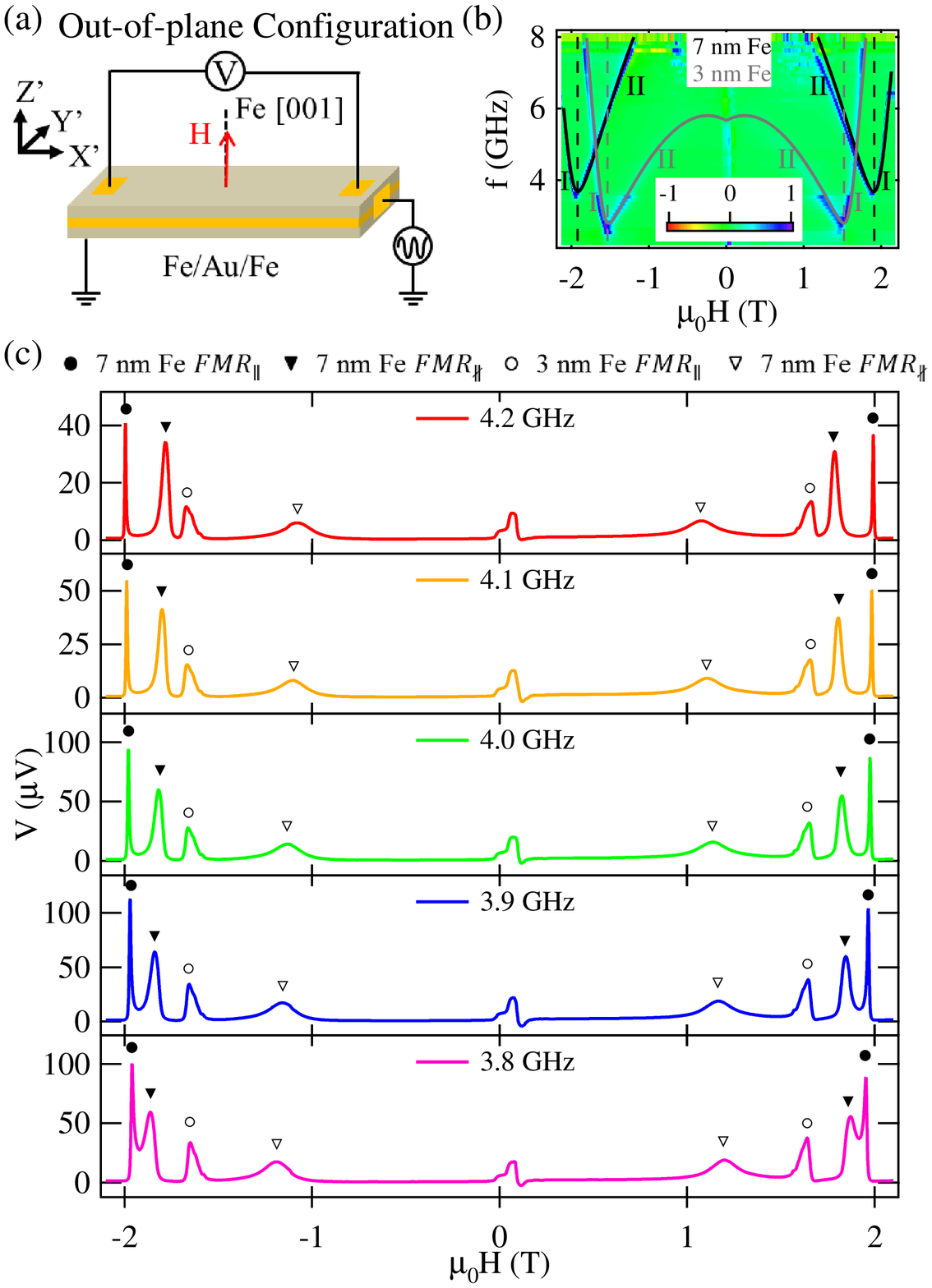}
\caption{$V_{SR}$ measurement with ${\mathbf H}$ pointing out of the film plane. (a) The sketch of the out-of-plane configuration measurement; (b) $\omega$-${\mathbf H}$ dispersive image plot, the grey solid line is the fitting curve of 3 nm Fe, and the black solid line is the fitting curve of 7 nm Fe; both dispersion curves have two brunches: brunch I is $FMR_{\parallel}$ brunch, and brunch II is $FMR_{\nparallel}$ brunch. (c) Typical curves in out-of-plane configuration, solid circles ($\bullet$) indicate peaks belong to $FMR_{\parallel }$ brunch in 7 nm Fe, solid triangles ($\blacktriangledown$) indicate peaks  belong to $FMR_{\nparallel }$ brunch in 7 nm Fe, hollow circles ($\circ$) indicate peaks belong to $FMR_{\parallel }$ brunch in 3 nm Fe, and hollow triangles ($\triangledown$) indicate peaks  belong to $FMR_{\nparallel }$ brunch in 3 nm Fe.}
\label{outplaneM}
\end{figure}

In addition to magneto anisotropy, shape anisotropy is also able to affect the relative angle between ${\mathbf M}$ and ${\mathbf H}$. Fig. \ref{outplaneM} shows the results when ${\mathbf H}$ is applied almost perpendicular to film plane, with Fig. \ref{outplaneM}(a) showing a sketch of out-of-plane configuration measurement with with ${\varphi _H}=0^\circ$ and ${\vartheta _H}\approx90^\circ$. In this configuration, when ${\mathbf H}$ is larger than the saturation field, ${\mathbf M}\parallel{\mathbf H}$, and when ${\mathbf H}$ is smaller than the saturation field, ${\mathbf M}\nparallel{\mathbf H}$. The relative angle between ${\mathbf M}$ and ${\mathbf H}$ is determined by the competition between the Zeeman energy and shape anisotropy energy. Fig. \ref{outplaneM}(b) shows $\omega$-${\mathbf H}$ dispersion plot, with normalized rectification voltage amplitude mapped into rainbow color scale as the indicator marks. We can identify the dispersion curve traced by the black solid line as originating from the 7 nm Fe layer and the curve traced by the grey solid line as originating from the 3 nm Fe layer. Both dispersion curves also have $FMR_{\parallel}$ brunch and $FMR_{\nparallel }$ brunch. Fig. \ref{outplaneM}(c) shows several typical curves measured in this configuration at various microwave frequencies between 3.8 GHz and 4.2 GHz. All the resonance peaks show the Lorenz line shape, and we describe the key features by the following Eqs. (\ref{e:V_outplane}):

\begin{subequations}
\label{e:V_outplane}
\begin{align}
At~{\varphi _H}=0^\circ, {\vartheta _H}\approx90^\circ:\nonumber\\
~~~~~~~~~~~~~~~~~~~~~~~~~~~~~&V(H)=V(-H)\label{e:V_outplaneA}\\
~~~~~~~~~~~~~~~~~~~~~~~~~~~~~&\frac{V_{Fe_{7}}}{|V_{Fe_{7}}|}=\frac{V_{Fe_{3}}}{|V_{Fe_{3}}|}\label{e:V_outplaneB}\\
~~~~~~~~~~~~~~~~~~~~~~~~~~~~~&\frac{V_{FMR_{\parallel}}}{|V_{FMR_{\parallel}}|}=\frac{V_{FMR_{\nparallel }}}{|V_{FMR_{\nparallel }}|}\label{e:V_outplaneC}
\end{align}
\end{subequations}

Equations (\ref{e:V_outplane}) are quite different from Eqs. (\ref{e:V_inplane}). Eq. (\ref{e:V_outplaneA}) shows the voltage signal keeps the same polarity when ${\mathbf H}$ reverses, which indicates the spin pumping and the inverse spin hall effect is ignorable in our measurement \cite{BaiLH}. Eq. (\ref{e:V_outplaneB}) shows the signal polarity in two Fe layers are the same and Eq. (\ref{e:V_outplaneC}) shows the signal polarity in $FMR_{\nparallel }$ brunch keeps the same as in $FMR_{\parallel }$ brunch. From Fig. (\ref{outplaneM}) (c), the resonance peaks in $FMR_{\nparallel}$ brunch is also much broader than in $FMR_{\parallel}$ brunch. Comparing Fig. (\ref{inplaneM}) and (\ref{outplaneM}), the SRE signal in $FMR_{\nparallel }$ brunch has the same behaviour as in $FMR_{\parallel }$ brunch when changing the measurement configuration and rf magnetic field direction. And comparing $V_{SR}$ in two brunches, the signal polarity is opposite in the in-plane configuration and keep the same in the out-of-plane configuration. 

So far in the literatures, SRE was systematically studied only in the configuration with ${\mathbf M}$ $\parallel$ ${\mathbf H}$, and the rectification voltage is described by a formula as a function of ${\mathbf H}$ \cite{SRETheory}. Since ${\mathbf M}$ and ${\mathbf H}$ are non-collinear in $FMR_{\nparallel }$ brunch, the conclusions in previous studies are not suitable here any more. However, the ${\mathbf M}$ alignment is always parallel to the effective field ${\mathbf H_{eff}}$ rather than ${\mathbf H}$. Thus, ${\mathbf H_{eff}}$ instead of ${\mathbf H}$ should be taken into account especially in ferromagnetic systems with strong anisotropy and demagnetization. ${\mathbf H_{eff}}$ is determined by the free energy ${\mathbf F}$ of the system. Considering the single crystal magnetic thin film in our case with Zeeman energy, magneto anisotropy energy, shape anisotropy and demagnetization energy, one can get the free energy ${\mathbf F}$ and the effective field ${\mathbf H_{eff}}$ as follows:

%
%
\begin{subequations}
\label{e:heffcal}
\begin{align}
&F = - {\mu _0}MH\left[\cos {\theta _H}\cos {\theta _M}\cos ({\varphi _M} - {\varphi _H}) + \sin {\theta _H}\sin {\theta _M}\right]\nonumber\\ &+ \frac{1}{2}{\mu _0}{M_{eff}}^2{\sin ^2}{\theta _M}
 + {K_{s,p}}{\cos ^2}{\theta _M}{\cos ^2}({\varphi _M} - {\varphi _{s,p}}) \nonumber\\&+ \frac{1}{4}{K_1}\left({\sin ^2}2{\theta _M} + {\cos ^4}{\theta _M}{\sin ^2}2{\varphi _M}\right) \label{e:heffcalG}\\
&H_{eff}^2 = {\left( {\frac{\omega }{\gamma }} \right)^2} \nonumber\\&= {\left. {\frac{{\mu _0^2}}{{{{\left( {{\mu _0}M{{\cos }}{\theta_M} } \right)}^2}}}\left[\frac{{{\partial ^2}F}}{{\partial {\theta_M ^2}}}\frac{{{\partial ^2}F}}{{\partial {\varphi_M ^2}}} - {{\left( {\frac{{{\partial ^2}F}}{{\partial {\varphi_M} \partial {\theta_M} }}} \right)}^2}\right]} \right|_{({\theta _M},{\varphi _M})}}\label{e:heffcalA}
\end{align}
\end{subequations}

\begin{figure}[configuration]
\includegraphics[width= 7.6 cm]{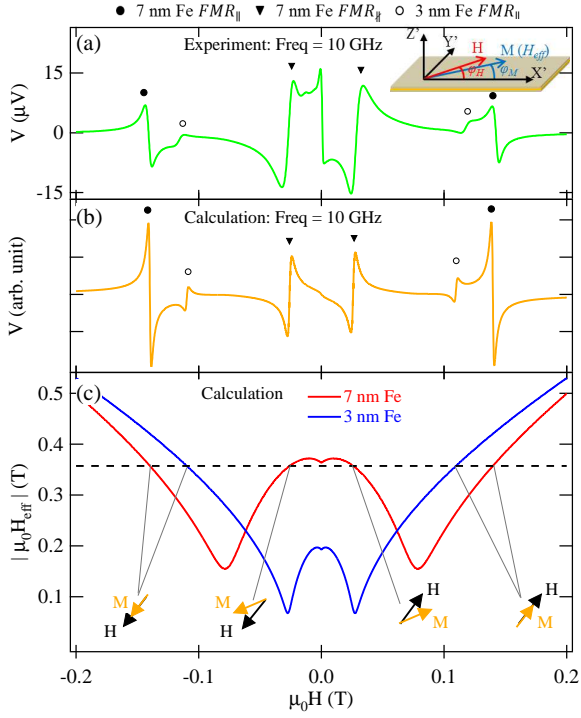}
\caption{Comparison of experiment and calculation results in the in-plane configuration. (a) Experiment curve with microwave frequency of 10 GHz, the inset shows non-collinear configuration of ${\mathbf M}$ and ${\mathbf H}$ in the in-plane configuration. (b) Calculation curve with microwave frequency fixed at 10 GHz, the result is agree with Eqs. (\ref{e:V_inplane}). Solid lines in (c) are calculated effective field as a function of the applied field of 7 nm Fe layer and 3 nm Fe layer in the in-plane configuration and the dashed line indicates the position of effective field which satisfies resonance condition with microwave frequency of 10 GHz.}
\label{theory_inplane}
\end{figure}

\begin{figure}[configuration]
\includegraphics[width= 7.6 cm]{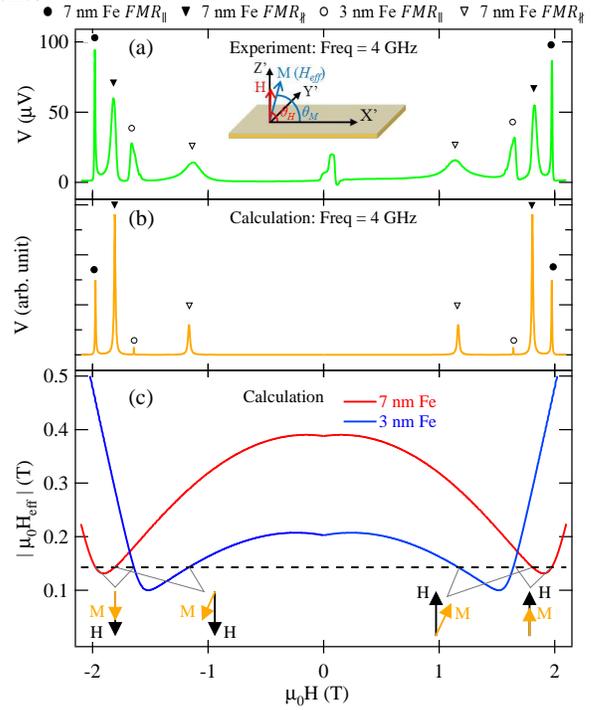}
\caption{Comparison of experiment and calculation results in the out-of-plane configuration. (a) Experiment curve with microwave frequency of 4 GHz, the inset shows non-collinear configuration of ${\mathbf M}$ and ${\mathbf H}$ in the out-of-plane configuration. (b) Calculation curve with microwave frequency fixed at 4 GHz, the result is agree with Eqs. (\ref{e:V_outplane}). Solid lines in (c) are calculated effective field as a function of the applied field of 7 nm Fe layer and 3 nm Fe layer in the out-of-plane configuration and the dashed line indicates the position of effective field which satisfies resonance condition with microwave frequency of 4 GHz.}
\label{theory_outplane}
\end{figure}

Here ${\varphi _M}$, ${\vartheta _M}$, ${\varphi _H}$ and ${\vartheta _H}$ are the angles of ${\mathbf M}$ and ${\mathbf H}$, as defined in insets of Fig. \ref{theory_inplane}(a) and Fig. \ref{theory_outplane}(a), ${\mu _0}$ is susceptibility in vacuum, ${\mathbf M_{eff}}$ is effective moment, $K_{s,p}$ is uniaxial anisotropy constant, ${\varphi _{s,p}}$ is the angle of easy axis of uniaxial anisotropy, and $K_{1}$ is four-fold anisotropy constant. Putting effective field calculated from Eq. (\ref{e:heffcal}) and the microwave magnetic field ${\mathbf h_{X'Y'Z'}} = \left( {0,{h_{Y'}}cos\left( \delta  \right){e^{i\omega t}},0} \right)$ in to LLG equation, we can get the dynamic magnetization ${\mathbf m}$. Here $\delta$ is the phase of the microwave field ${\mathbf h}$, and in our system we define $\delta=0$ in 7 nm Fe and $\delta=\pi$ in 3 nm Fe. Spin rectification voltage is described as ${V_{SR}} = \left\langle {j * \Delta R} \right\rangle $, here $j$ is microwave current in the system, and $ \Delta R\propto Re(m) $ is resistance variation within the system  due to AMR and spin procession. Thus we can derive the SRE in the in-plane configuration:
\begin{eqnarray}
V_{SR} = A\ast{\mathop{\rm Re}\nolimits} ({\chi _T}){h_{Y'}}cos\left( \varphi _M+\delta  \right)\sin \left( {2{\varphi _M}} \right)\label{inplane}
\end{eqnarray}
and the SRE in the out-of-plane configuration:
\begin{eqnarray}
V_{SR} =   A\ast{\mathop{\rm Re}\nolimits} \left( {{\chi _L}} \right){h_{Y'}}\sin \left( {2{\theta _M}} \right)\label{outplane}
\end{eqnarray}
with
\begin{subequations}
\begin{align}
&A=-\frac{{{j_{x'}}\Delta R}}{2M}\nonumber\\
&{\mathop{\rm Re}\nolimits} \left( {{\chi _L}} \right) =  - \frac{{{\omega _M}{\omega _{{H_{eff}}}}\left( {\omega _{{H_{eff}}}^2 - {\omega ^2}} \right)}}{{{{\left( {\omega _{{H_{eff}}}^2 - {\omega ^2}} \right)}^2} + 4\omega _{{H_{eff}}}^2{\alpha ^2}{\omega ^2}}}\nonumber\\
&{\mathop{\rm Re}\nolimits} \left( {{\chi _T}} \right) = \frac{{2\alpha {\omega ^2}{\omega _M}{\omega _{{H_{eff}}}}}}{{{{\left( {{\omega ^2}_{{H_{eff}}} - {\omega ^2}} \right)}^2} + 4{\omega _{{H_{eff}}}^2}{\alpha ^2}{\omega ^2}}}\nonumber
\end{align}
\end{subequations}

Here $j_{x'}$ is the microwave current amplitude, ${\mathop{\rm Re}\nolimits} \left( {{\chi _L}} \right)$ and ${\mathop{\rm Re}\nolimits} \left( {{\chi _T}} \right)$ are respectively the real parts of diagonal and non-diagonal elements of dynamic susceptibility tensor, $\omega$ is the applied microwave frequency, $\omega_{M}=\gamma M$, $\omega_{H_{eff} }=\gamma H_{eff}$, $\gamma$ is gyromagnetic ratio, and $\alpha$ is damping constant. As shown in Eq. (\ref{inplane}) and (\ref{outplane}), $V_{SR}$ is a function of the effective field ${\mathbf H_{eff}}$ instead of the applied field ${\mathbf H}$, thus $V_{SR}$ cannot be fitted by a simple formula. To analysis the SRE, we first get ${\varphi _M}$ and ${\vartheta _M}$ as functions of ${\mathbf H}$ by minimizing system Free energy F as shown in Eq. (\ref{e:heffcalG}), then calculate effective field ${\mathbf H_{eff}}$ by Eq. (\ref{e:heffcalA}), and finally calculate $V_{SR}$ by Eq. (\ref{inplane}) and (\ref{outplane}).


Figure \ref{theory_inplane} shows the comparison between calculation and experimental results in the in-plane configuration. Fig. \ref{theory_inplane}(a) is a typical experimental curve measured with the microwave frequency of 10 GHz, and (b) shows the calculation curve with the microwave frequency fixed at 10 GHz, the effective field ${\mathbf H_{eff}}$ as a function of ${\mathbf H}$ in the in-plane configuration is shown in (c). Here we use ${\varphi _H=44.6^\circ}$ and ${\vartheta _H=0^\circ}$ for the in-plane configuration, $\mu_{0}H_{1}=0.073 ~T$, $\mu_{0}M_{eff}=1.7 ~T$ for 7 nm Fe, and $\mu_{0}H_{1}=0.026 ~T$, $\mu_{0}M_{eff}=1.4 ~T$ for 3 nm Fe. These parameters are all determined by the dispersion curves in Fig. \ref{inplaneM}(b). And we use $\alpha=0.006$ calculated from line width, and set $A_{Fe_7} = 5 \times A_{Fe_3}$ to best represent the experimental conditions. The calculation results agree well with experimental results. From Eq. (\ref{inplane}), $V_{SR}$ is determined by the real part of diagonal elements of dynamic susceptibility tensor ${\mathop{\rm Re}\nolimits} \left( {{\chi _L}} \right)$ which is anti Lorentz line shape, so $V_{SR}$ is anti Lorentz line shape as shown in Fig. \ref{theory_inplane}(b) and confines with experimental result. Since $V_{SR}\propto\cos\left( \varphi _M+\delta  \right)\sin \left( {2{\varphi _M}} \right)$, and when ${\mathbf H}$ reverses, the ${\mathbf H_{eff}}$ and ${\mathbf M}$ will reverse, which corresponds to ${\varphi _M}+\pi$ and ${\vartheta _M}+\pi$, $V_{SR}$ will change its polarity when ${\mathbf H}$ reverses as shown in Fig. \ref{theory_inplane}(b) and confines with Eq. (\ref{e:V_inplaneA}). And $V_{SR}$ has opposite polarity in 7 nm Fe layer and 3 nm Fe layer, as shown in Fig. \ref{theory_inplane}(b) and confines with Eq. (\ref{e:V_inplaneB}), because in 7 nm Fe layer and 3 nm Fe layer, the phase $\delta$ of rf magnetic field ${\mathbf h}$ has a difference of $\pi$. As shown in Fig. \ref{theory_inplane}(c), the ${\mathbf H_{eff}}$ will increase as ${\mathbf H}$ increases when ${\mathbf M}$ and ${\mathbf H}$ are collinear, which means spin procession is in-phase when $H>H_{FMR}$ and out-of-phase when $H<H_{FMR}$ \cite{phase}, while the ${\mathbf H_{eff}}$ will decrease as ${\mathbf H}$ increases when ${\mathbf M}$ and ${\mathbf H}$ are non-collinear, which means spin procession is in-phase when $H<H_{FMR}$ and out-of-phase when $H>H_{FMR}$ \cite{phase}. Near the resonance position as indicated by dashed line in Fig. \ref{theory_inplane}(c),  $\left(H-H_{FMR}\right)/\left(H_{eff}-H_{FMR}\right)>0$ in $FMR_{\parallel}$ brunch, while $\left(H-H_{FMR}\right)/\left(H_{eff}-H_{FMR}\right)<0$ in $FMR_{\nparallel}$ brunch. And Since the sign of $V_{SR}$ is determined by ${\left( {\omega _{{H_{eff}}}^2 - {\omega ^2}} \right)}$, $V_{SR}$ has the opposite polarity in $FMR_{\parallel}$ brunch and $FMR_{\nparallel}$ brunch when $V_{SR}$ is plotted as a function of ${\mathbf H}$, as shown in Fig. \ref{theory_inplane}(b) and confines with Eq. (\ref{e:V_inplaneC}).

Our theory also works in the out-of-plane configuration. Fig. \ref{theory_outplane} shows the comparison between calculation and experimental results in the out-of-plane configuration. Fig. \ref{theory_outplane}(a) is a typical experimental curve measured with the microwave frequency of 4 GHz, and (b) shows the calculation curve with the microwave frequency fixed at 4 GHz, the effective field ${\mathbf H_{eff}}$ as a function of ${\mathbf H}$ in the out-of-plane configuration is shown in (c). In calculation, we use ${\varphi _H=0^\circ}$ and ${\vartheta _H=89.4^\circ}$ for the out-of-plane configuration, and keep the other parameters the same as those used in the in-plane configuration. From Eq. (\ref{outplane}), $V_{SR}$ is determined by the real part of non-diagonal elements of dynamic susceptibility tensor ${\mathop{\rm Re}\nolimits} \left( {{\chi _T}} \right)$ which is Lorentz line shape, as shown in Fig. (\ref{theory_outplane})(b), and confines with experimental results. Since $V_{SR}\propto\sin \left( {2{\theta_M}} \right)$, $V_{SR}$ keeps the same polarity when ${\mathbf H}$ reverses (confines with Eq. (\ref{e:V_outplaneA})), and keeps the same polarity in 7 nm Fe and 3 nm Fe layer (confines with Eq. (\ref{e:V_outplaneB})). And Since the sign of $V_{SR}$ is determined by ${\omega _{{H_{eff}}}}$, $V_{SR}$ polarity keeps the same in $FMR_{\parallel}$ brunch and $FMR_{\nparallel}$ brunch. 

The calculation and the experimental results of the SRE in 7 nm Fe layer are listed in Table \ref{table}. Our theory well describes the line shape and polarity of the SRE in the general configuration with ${\mathbf M}$ and ${\mathbf H}$. Also our theory confirms the broaden of linewidth $\Delta H$ when ${\mathbf M}$ and ${\mathbf H}$ are non-collinear qualitatively. However, the broaden of linewidth in experiment is larger, and the quantitative analysis still needs further discussions.

\begin{widetext}
\begin{center}
\begin{table}
\begin{tabular}{ l        l        l        l        l        l }
\hline\hline
 & Measurement Configuration & Line shape & polarity & $\mu_{0}H_{FMR}~(T)$ & $\mu_{0}\Delta H~(T)$ \\ \hline
\multirow{4}{*}{M $\parallel$ H}~~~~ & \multirow{2}{*}{${\varphi _H=44.6^\circ}, {\vartheta _H=0^\circ}, f=10$ GHz}~~~~ & Anti-Lorentz$_{ (Exp)}$ ~~~~ & ${-_{( Exp)}}$ ~~~~    & ${0.14_{( Exp)}}$   ~~~~  & ${0.0053_{( Exp)}}$\\ 
 & & Anti-Lorentz$_{ (Cal)}$~~~~ & $-_{( Cal)}$~~~~ & $0.14_{( Cal)}$~~~~ & $0.0024_{( Cal)}$\\
 & \multirow{2}{*}{${\varphi _H=0^\circ}, {\vartheta _H=89.4^\circ}, f=4$ GHz}~~~~     & Lorentz$_{ (Exp)}$ ~~~~    & $+_{(Exp)}$ ~~~~    & $1.96_{(Exp)}$~~~~     & $0.0095_{(Exp)}$ \\
 & & Lorentz$_{ (Cal)}$~~~~ & $+_{(Cal)}$~~~~ & $1.97_{(Cal)}$~~~~ & $0.0072_{(Cal)}$\\
\multirow{4}{*}{M $\nparallel$ H}~~~~ & \multirow{2}{*}{${\varphi _H=44.6^\circ}, {\vartheta _H=0^\circ}, f=10$ GHz} ~~~~    & Anti-Lorentz$_{ (Exp)}$~~~~  & ${+_{( Exp)}}$~~~~     & ${0.027_{( Exp)}}$~~~~     & ${0.010_{( Exp)}}$\\ 
 & & Anti-Lorentz$_{ (Cal)}$~~~~ & +$_{( Cal)}$~~~~ & $0.026_{( Cal)}$~~~~ & $0.0036_{( Cal)}$\\
 & \multirow{2}{*}{${\varphi _H=0^\circ}, {\vartheta _H=89.4^\circ}, f=4$ GHz}~~~~     & Lorentz$_{ (Exp)}$ ~~~~    & $+_{(Exp)}$ ~~~~    & $1.79_{(Exp)}$~~~~     & $0.054_{(Exp)}$ \\
 & & Lorentz$_{ (Cal)}$ ~~~~& $+_{(Cal)}$~~~~ & $1.80_{(Cal)}$~~~~ & $0.013_{(Cal)}$\\
\hline\hline
\end{tabular}
\caption{The calculation and the experimental results of the SRE in the ${\mathbf M}$ $\parallel$ ${\mathbf H}$ and the ${\mathbf M}$ $\nparallel$ ${\mathbf H}$ configuration in 7 nm Fe layer with different measurement geometry. The positive polarity of the SRE is defined as ${V_{SR}/\vert V_{SR}\vert>0}$ when $H<H_{FMR}$. The subscript Exp indicates the result is extracted from the experimental data, and the subscript Cal indicates the result is extracted from the calculation data.}
\label{table}
\end{table}
\end{center}
\end{widetext}

In conclusion, we studied Spin Rectification Effect in an epitaxial Fe/Au/Fe tri-layer system with strong magneto anisotropy and shape anisotropy. In addition to the SRE when ${\mathbf M}$ and ${\mathbf H}$ are collinear, we study $V_{SR}$ for the case where M and H are non-collinear. The different behaviour of $V_{SR}$ in different configuration of ${\mathbf M}$ and ${\mathbf H}$ are due to the different relationship of the ${\mathbf H_{eff}}$ depending on ${\mathbf H}$. By considering ${\mathbf H_{eff}}$ instead of ${\mathbf H}$ in ferromagnetic system, we extend the SRE theory for all ${\mathbf M}$ and ${\mathbf H}$ configurations in different measurement configuration. These equations will help further understanding of spin transport in ferromagnetic systems, especially when ${\mathbf M}$ is not parallel to ${\mathbf H}$.

{\bf Acknowledgments}
This project was supported by the National Key Basic Research Program (Grants No. 2015CB921401 and No. 2011CB921801), National Science Foundation (Grants No. 11274074, 11434003, 11474066 and 11429401) of China, and NSERC grands. The authors thank J. X. Li from Fudan University, Z. H. Zhang, B. M. Yao, L. Fu and Y. S. Gui from University of Manitoba, and X. L. Fan from Lanzhou University.
%

%

%
\end{document}